\documentclass[12pt,preprint]{aastex}

\usepackage{amsmath,natbib}
\usepackage{rotating}
\usepackage{graphicx}

\def\eg{{\it e.g.,\,}}

\def\half{\frac{1}{2}}
\def\spose#1{\hbox to 0pt{#1\hss}}
\def\lta{\mathrel{\spose{\lower 3pt\hbox{$\mathchar"218$}}
     \raise 2.0pt\hbox{$\mathchar"13C$}}}
\def\gta{\mathrel{\spose{\lower 3pt\hbox{$\mathchar"218$}}
     \raise 2.0pt\hbox{$\mathchar"13E$}}}

\newcommand{\COSMOMC}{\textsc{CosmoMC}}
\newcommand{\CBI}{CBI}
\newcommand{\Mpc}{\text{Mpc}}
                   
\slugcomment{submitted to ApJ}

\shorttitle{CBI Polarization Implications}
\shortauthors{Sievers et al.}

\begin{document}

\title{Implications of the Cosmic Background Imager
  Polarization Data}

\author{J.~L.~Sievers\altaffilmark{1}, C.~Achermann\altaffilmark{2},
J.~R.~Bond\altaffilmark{1}, L.~Bronfman\altaffilmark{3},
R.~Bustos\altaffilmark{2},
C.~R.~Contaldi\altaffilmark{1}$^,$\altaffilmark{4},
C.~Dickinson\altaffilmark{5}, P.~G.~Ferreira\altaffilmark{6},
M.~E.~Jones\altaffilmark{6}, A.~M.~Lewis\altaffilmark{1},
B.~S.~Mason\altaffilmark{7}, J.~May\altaffilmark{3}, 
S.~T.~Myers\altaffilmark{8}, S.~Padin\altaffilmark{9},
T.~J.~Pearson\altaffilmark{5}, M.~Pospieszalski\altaffilmark{10},
A.~C.~S.~Readhead\altaffilmark{5}, R.~Reeves\altaffilmark{2},
A.~C.~Taylor\altaffilmark{6} and S.~Torres\altaffilmark{2}}

\altaffiltext{1}{Canadian Institute for Theoretical
Astrophysics, University of Toronto, ON M5S 3H8, Canada} 
\altaffiltext{2}{Departamento de Ingenier{\'\i}a El{\'e}ctrica, Universidad de Concepci{\'o}n, Concepci{\'o}n, Chile} 
\altaffiltext{3}{Departamento de Astronom{\'\i}a, Universidad de Chile,  Santiago, Chile}
\altaffiltext{4}{Department of Physics, Imperial College, London, UK}
\altaffiltext{5}{Owens Valley Radio Observatory, California Institute of
Technology, Pasadena, CA}
\altaffiltext{6}{Astrophysics, Oxford University, Keble Road, Oxford OX1 3RH, UK} 
\altaffiltext{7}{National Radio Astronomy Observatory, Green Bank, WV 24944} 
\altaffiltext{8}{National Radio Astronomy Observatory, Socorro, NM 87801} 
\altaffiltext{9}{Kavli Institute for Cosmological Physics, Department of Astronomy and Astrophysics, 
University of Chicago, Chicago, IL 60637}
\altaffiltext{10}{National Radio Astronomy Observatory, 520 Edgemont Road, Charlottesville, VA 22903}

\begin{abstract}
We present new measurements of the power spectra of the E-mode of CMB
polarization, the temperature T, the cross-correlation of E and T, and
upper limits on the B-mode from 2.5 years of dedicated Cosmic
Background Imager (CBI) 
observations. Both raw maps and optimal signal images in the $uv$-plane
and the sky plane show strong detections of the E-mode (11.7$\sigma$ for 
the EE power spectrum overall) and no detection
of the B-mode. The power spectra are used to constrain parameters of
the flat tilted adiabatic $\Lambda$CDM models: those determined from
EE and TE bandpowers agree with those from TT, a powerful consistency
check. There is little tolerance for shifting polarization peaks from
the TT-forecast locations, as measured by the angular sound crossing
scale $\theta = 100 /\ell_s = 1.03 \pm 0.02$ from EE and TE {\it cf.}
$1.044 \pm 0.005$ with the TT data included. The scope for extra
out-of-phase peaks from subdominant isocurvature modes is also
curtailed. The EE and TE measurements of CBI, DASI and BOOMERANG are
mutually consistent, and, taken together rather than singly, give
enhanced leverage for these tests.

\end{abstract}


\keywords{cosmology, cosmic microwave background, polarization}

\section{Introduction}

Polarization of the cosmic microwave background (CMB) at the $\sim
10\%$ level has been forecast for decades \eg \cite{BE84}, but only
after a long experimental struggle was it detected and firmly
established, from measurements by the DASI 
\citep{Kovac:2002fg}, WMAP \citep{Kogut:2003et}, CBI \citep{paper8},
CAPMAP~\citep{capmap04} and
Boomerang~\citep{Masi:2005,Montroy:2005,Piacentini:2005} experiments.
The CBI \citep{Padin02, paper8} is a 13-element interferometer located
at 5080 meters in the Chilean Andes operating in ten 1-GHz bands from
26 GHz to 36 GHz. The CBI has been observing the polarization of the
CMB since its inception in late 1999, and since 2002 September has
been operating in a polarization optimized configuration with 42
polarization-sensitive baselines.  The first
\CBI\ polarization limits \citep{Cart05} used only 12
polarization-sensitive baselines. 
As part of the polarization optimization, we adopted the achromatic
polarizers designed by Kovac and described in \cite{Kovac:2002fg}.  The first
CBI detections in the 
polarization-optimized configuration included data taken from 2002
September to 2004 May \citep{paper8}. In this paper, we present and
analyze the implications of CBI polarization for our complete 2002
September to 2005 April polarization dataset. This represents a 54\%
increase in integration time over the results reported earlier
\citep{paper8}.  

In \S~\ref{sec:data}, we present an abbreviated description of the
data and its analysis that leads to the compression of the data on to
maps in $\ell$-space, a natural space for interferometry of the CMB, and
further compression on to power spectra. A detailed description of the
experiment, our analysis procedure and results of data quality tests
will be given in Myers et al.\ (2006, in preparation). 
We use our improved EE and TE power
spectra to test consistency of cosmological parameters with results
forecast from TT, for minimal inflation-motivated tilted $\Lambda$CDM
models in \S~\ref{sec:ad_params} and hybrid models with an additional
subdominant isocurvature component in \S~\ref{sec:iso_params}. Special
attention is paid to overall amplitudes and pattern-shifting
parameters in \S~\ref{sec:ad_pattern} and \S~\ref{sec:iso_pattern}.

\section{Processing of the CBI Polarization Data}
\label{sec:data}

\subsection{The Polarization Data and the CBI pipeline}
\label{sec:pipeline}

The CBI instrument is described by \citet{Padin02} and the observing
and data reduction procedures used for CMB polarization studies are
described in \cite{paper8}. Observations were made of four mosaic
fields, labeled $02^{\rm h}$, $08^{\rm h}$, $14^{\rm h}$, and $20^{\rm
h}$ by right ascension, each having roughly equal observing time. The
$20^{\rm h}$ field was a deep $4.5^\circ \times 0.75^\circ$ $20^{\rm
h}$ strip of $6$ pointings separated by $45^\prime$. The {\it fwhm} of
the CBI primary beam is $45^\prime$. The other three mosaics were $6
\times 6$ pointings each covering a $4.5^\circ \times 4.5^\circ$
square. The CBI recorded visibilities on 78 baselines (antenna pairs).
We note that the data in \citet{paper8} used only 12 of the 13
antennas, or 66 baselines, due to a software error, which has now been
corrected.  This error resulted in a ~15\% loss in sensitivity, but
did not bias results.  The data were calibrated by reference to
standard sources with an uncertainty of 1.3\% in flux density
\citep{paper8}.  Six of the antennas were set to receive right
circular polarization ($R$) and seven left ($L$), so each visibility
measurement represents one of the four polarization products
$RR,RL,LR,LL$. The copolar products $RR$ and $LL$ are sensitive to
total intensity or brightness temperature T (under the assumption that
the circular polarization V is zero, as expected), while $RL$ and $LR$
are sensitive to linear polarization which can be divided into
grad-mode E and curl-mode B components.  In the small angle
approximation, where the celestial sphere can be described by a
tangent plane (``image'' or ``sky'' plane), the angular spherical
harmonic multipoles defining the CMB radiation field, labeled by
($\ell,m$), become Cartesian components of $\ell$, $(\ell_u,\ell_v)$.
In this same approximation, interferometer visibilities sample the
Fourier transform of the sky brightness, with $(u,v)$ as the conjugate
variables to angles on the celestial plane; this ``$uv$-space'' is
related to ``$\ell$-space'' by $(\ell_u,\ell_v)=2\pi$(u,v).  Each
interferometer baseline is therefore sensitive to emission on angular
scales centered on spherical harmonic multipole $\ell =2\pi x$ where
$x$ is the antenna separation in wavelengths.

The visibilities are processed by convolution with an $\ell$-space
gathering kernel to produce gridded
estimators $\Delta_{iQ}$ for polarizations $i=$T,E,B and covariance
matrix elements $C_{N(ii^\prime)QQ^\prime}$,
$C_{T(ii^\prime)QQ^\prime}$, $C_{P(ii^\prime)QQ^\prime}$ for
(Gaussian) instrumental noise, (Gaussian) CMB signals, and projection
templates associated with point sources and ground spillover
\citep{Myers03,paper8}.  This gridding compresses the $\sim
10^7$ visibilities to $\sim 10^4$ $\ell$-space ``pixels'' (labeled by
$Q$) without loss of essential information.

For calculation of angular power spectra, the gridded estimators and
covariance matrices are passed to a maximum likelihood procedure which
estimates the CMB polarization bandpowers $q_{Xb}$ in band $b$,
$X$=$(ii^\prime)$=TT,EE,TE,BB, associated noise bandpowers $q_{NXb}$,
a Fisher (or likelihood curvature) matrix $F_{Xb,X^\prime b^\prime}$
whose inverse encodes the variance around the maximum likelihood, and
$\ell$-space window functions $W_{Xb\ell}$.  The $q_{Xb}$ are defined
by the expansion of the spectra ${\cal C}^{X}_\ell = \sum_b q_{Xb}
{\cal C}^{X}_{\ell b}$. (Here ${\cal C}^{X}_\ell \equiv \ell (\ell +1)
\langle a_{i\ell m}a_{i^\prime\ell m}^*\rangle /2\pi$, where $a_{i\ell
m}$ denotes the multipole coefficients of the signal $s_i$,
$i=$T,E,B.)  To determine bandpowers, we use (theory-blind) flat
${\cal C}^{X}_{\ell b}$ shapes with top hat binning (unity inside and
zero outside of the band). The window functions from the pipeline
convert theory power spectra ${\cal C}^{X}_\ell$ into bandpowers
$q_{Xb} = \sum_\ell {\cal C}^{X}_\ell W_{Xb\ell}\, (\ell+1/2)/(\ell
(\ell +1))$ to compare with the observed ones.

\subsection{The Power Spectra}
\label{sec:bandpows}

\begin{figure*}[!t]
\centering
\includegraphics[width=0.95\textwidth]{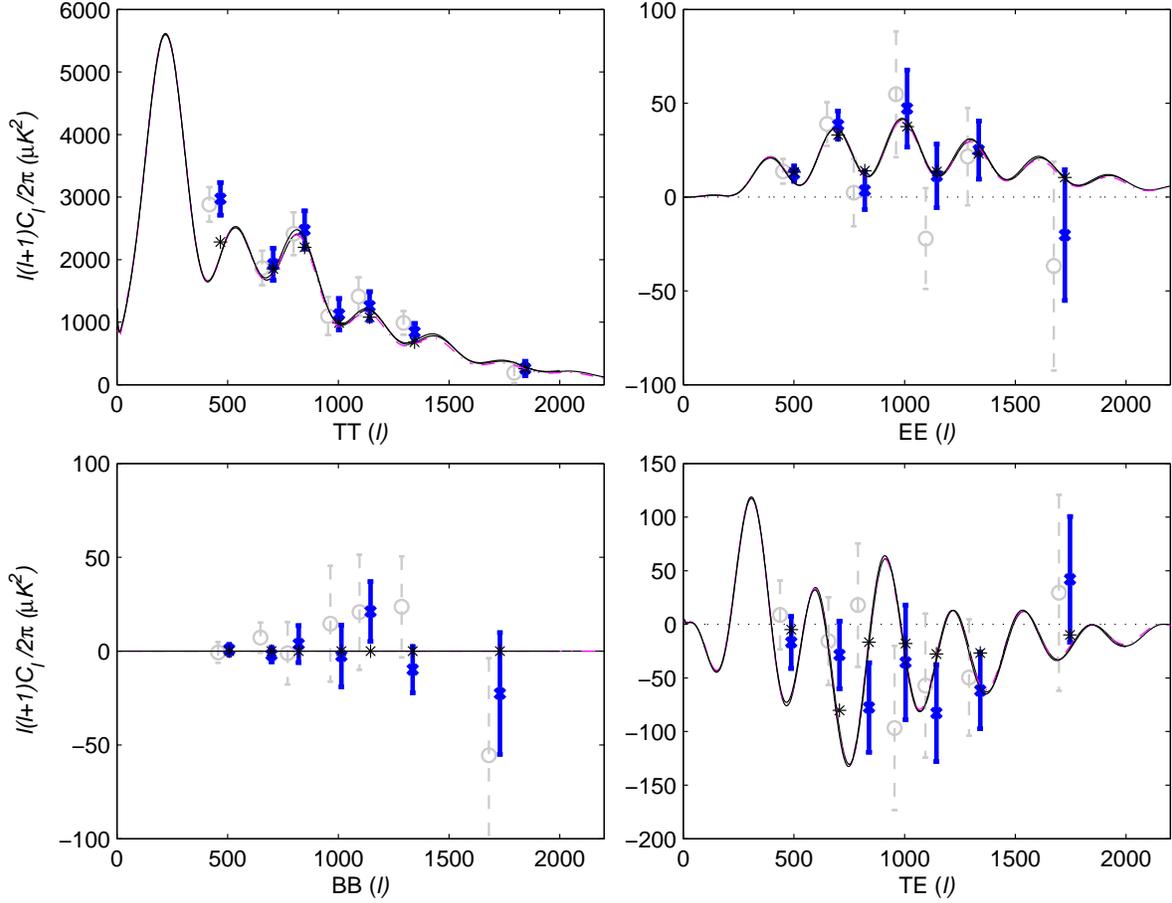}
\caption{The blue crosses show power spectra ${\cal C}_{\ell}^{X}$ as a
function of multipole $\ell$ derived from the CBI 2002 September -
2005 April data, for the total intensity TT, the grad polarization EE,
the curl polarization BB, and cross-correlation TE.  The {\it magenta dashed
curve} is a flat power-law $\Lambda$CDM model which best-fits the
{WMAP}, CBI, and ACBAR CMB data \citep{Spergel03}. It is very nearly
the black fiducial ${\cal C}_{\ell}^{X({\rm s})}$ which best-fits
WMAP1+CBI+DASI+B03 TT+EE+TE, whose parameters are given in
Table~\ref{tab:base}. The black asterisks show the expected values of
the bandpowers calculated from the fiducial model using the window functions.
The $\chi^2$ values of the data relative to the fiducial
model are 13.1, 2.25, 2.90, and 8.43 for TT, EE, BB, and TE for 7
degrees of freedom.  As expected, the BB spectrum is consistent with zero,
with a single-band amplitude of $0.2 \pm 1.6 \, \mu\rm{K}^2$, and a 95\%
upper limit of $3.76\,  \mu\rm{K}^2$.  (The grey circles show the
results of \citet{paper8} for comparison. The shrinking in error bars
is primarily due to the 54\% increase in the data.)  The CBI's
low-$\ell$ ($\ell \lta 360$) response is set by the details of the
sidelobes of the primary beam, which are difficult to measure; analysis
with finer bins shows that the apparent discrepancy in the first TT
bin plotted here is confined to $\ell < 360$ where the wings of the
primary beam are picking up the first Doppler peak.  Consequently we
neglect the $\ell<360$ TT from CBI in further analysis.  The data have been
offset in $\ell$ for clarity.}
\label{fig:7BinSpec}
\end{figure*}

\begin{figure*}[!t]
 \centering
\includegraphics[height=0.65\textwidth, angle=0]{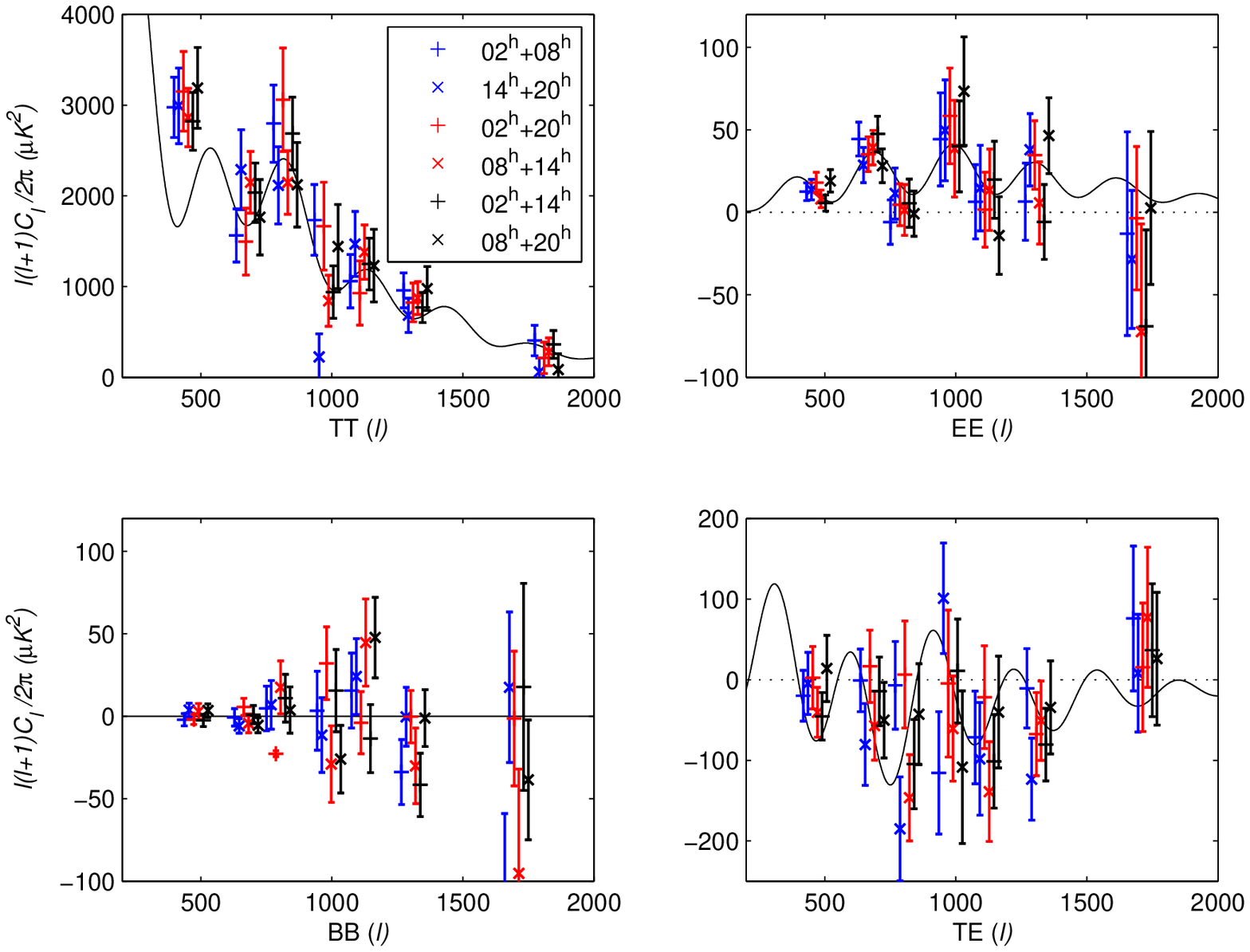}
\caption{The power spectra for the 6 distinct pairs of CBI fields,
  each denoted by the colors and symbols indicated in the TT
  panel. For example. the blue +'s show the $02^h+08^h$ fields and the
  blue x's show the $14^h+20^h$ fields. The pairs of spectra plotted
  with the same color are disjoint data sets and hence have meaningful
  $\chi^2$ values. For EE, the $\chi^2$ values of the disjoint subsets
  are 2.98 ($02^h+08^h$ vs. $14^h+20^h$), 8.55 ($02^h+14^h$
  vs. $08^h+20^h$), and 3.67 ($02^h+20^h$ vs. $08^h+14^h$) for 7
  degrees of freedom.  This confirms that the remarkably low $\chi^2$
  (2.25 for 7 dof) of the joint EE spectrum relative to the fiducial
  model is due to random chance.  The points have been spread in
  $\ell$ about the bin center for clarity.}

\label{fig:field_splits}
\end{figure*}

Fig.~\ref{fig:7BinSpec} shows the CBI maximum likelihood bandpowers
and their inverse-Fisher-matrix errors. The numerical values are given
in Table~\ref{tab:7bandpows}.  To minimize band-to-band correlations
only 7 bands are shown for EE, but we use about twice as many bands
for our cosmic parameter analyses. A nonlinear transformation to a
Gaussian in the ``offset lognormal'' combination $\ln( q_{Xb}+
q_{NXb})$ is used to give a more accurate representation of the
bandpower likelihood surface \citep{BJK98,Sievers03}.
Fig.~\ref{fig:field_splits} shows the spectra of pairs of fields of
the CBI data (which have better sensitivity than individual
fields).  These demonstrate that the remarkably good agreement between
the CBI EE spectrum and the fiducial model for EE evident in
Fig.~\ref{fig:7BinSpec} is due to random chance.

\begin{table*}
\centering
\space
\caption{CBI 7-Band Power Spectra (${\cal C}_\ell$ in $\mu\rm{K}^2)$.
\label{tab:7bandpows}}
\scriptsize
\begin{tabular}{|c||c|c|c|c|}
\hline\hline
& & & &  \\
 $\ell$-range  & TT &  EE & TE & BB \\
& & & &   \\				
\hline
 & & & & \\

$ <  600      $  & $   2971 \pm  260 $ & $   12.5 \pm  3.9 $ & $  -16.9 \pm 24.4 $ & $    0.8 \pm  2.8 $ \\
$  600 -  750 $  & $   1925 \pm  252 $ & $   38.3 \pm  7.6 $ & $  -28.5 \pm 31.3 $ & $   -1.9 \pm  4.0 $ \\
$  750 -  900 $  & $   2475 \pm  304 $ & $    3.6 \pm  9.9 $ & $  -77.6 \pm 41.9 $ & $    3.8 \pm 11.0 $ \\
$  900 - 1050 $  & $   1126 \pm  248 $ & $   47.2 \pm 20.6 $ & $  -35.5 \pm 53.7 $ & $   -2.6 \pm 17.2 $ \\
$ 1050 - 1200 $  & $   1256 \pm  239 $ & $   11.3 \pm 17.8 $ & $  -82.8 \pm 45.7 $ & $   21.1 \pm 16.1 $ \\
$ 1200 - 1500 $  & $    841 \pm  137 $ & $   25.0 \pm 15.2 $ & $  -61.7 \pm 35.7 $ & $   -9.9 \pm 12.0 $ \\
$ > 1500      $  & $    256 \pm  118 $ & $  -20.3 \pm 35.4 $ & $   42.0 \pm 59.6 $ & $  -22.6 \pm 31.6 $ \\

\hline\hline

\end{tabular}

\end{table*}

If the overall amplitude of $C_P$ is set to a large value then the
``nuisance'' modes represented in the construction of the matrix are
projected out. This is essential for ground subtraction and point
source projections for T.  As discussed in \cite{paper8}, where we
project out the brightest $\sim$20\% of NVSS sources in E and B, we
see no evidence for any polarized point sources in the CBI data.  In
particular, for CBI the EE spectrum is strongly detected while the BB
spectrum is consistent with zero, whereas uncorrelated polarized point sources
give rise to (roughly) equal amounts of power in EE and BB.  Neither
polarization power spectrum rises like $\ell^2$ as would be expected from an
appreciable point source signal.  
The spectra with and without the brightest $\sim$20\% of NVSS sources
projected out in polarization are very similar, with no systematic
trend in the differences.
Finally, there are no sources
visible in the CBI polarization maps.  Consequently, in this work, we
do not project out any sources in polarization.  
The nature of the 30~GHz polarized radio source population is currently
poorly known, but some models have been devised based on extrapolation
from lower frequency surveys.  These generally indicate that contamination
of the EE and BB power spectra should be negligible for the CBI $\ell$
range, observing frequency, and sensitivity levels, \eg\ \cite{Tucci:2003wd}
Further discussion
of this issue will be presented in Myers et al.\ (2006, in 
preparation).  

\subsection{Raw Maps and Signal Reconstructed Maps}
\label{sec:maps}

\begin{figure*}[!t]
 \centering
\includegraphics[height=0.95\textwidth, angle=0]{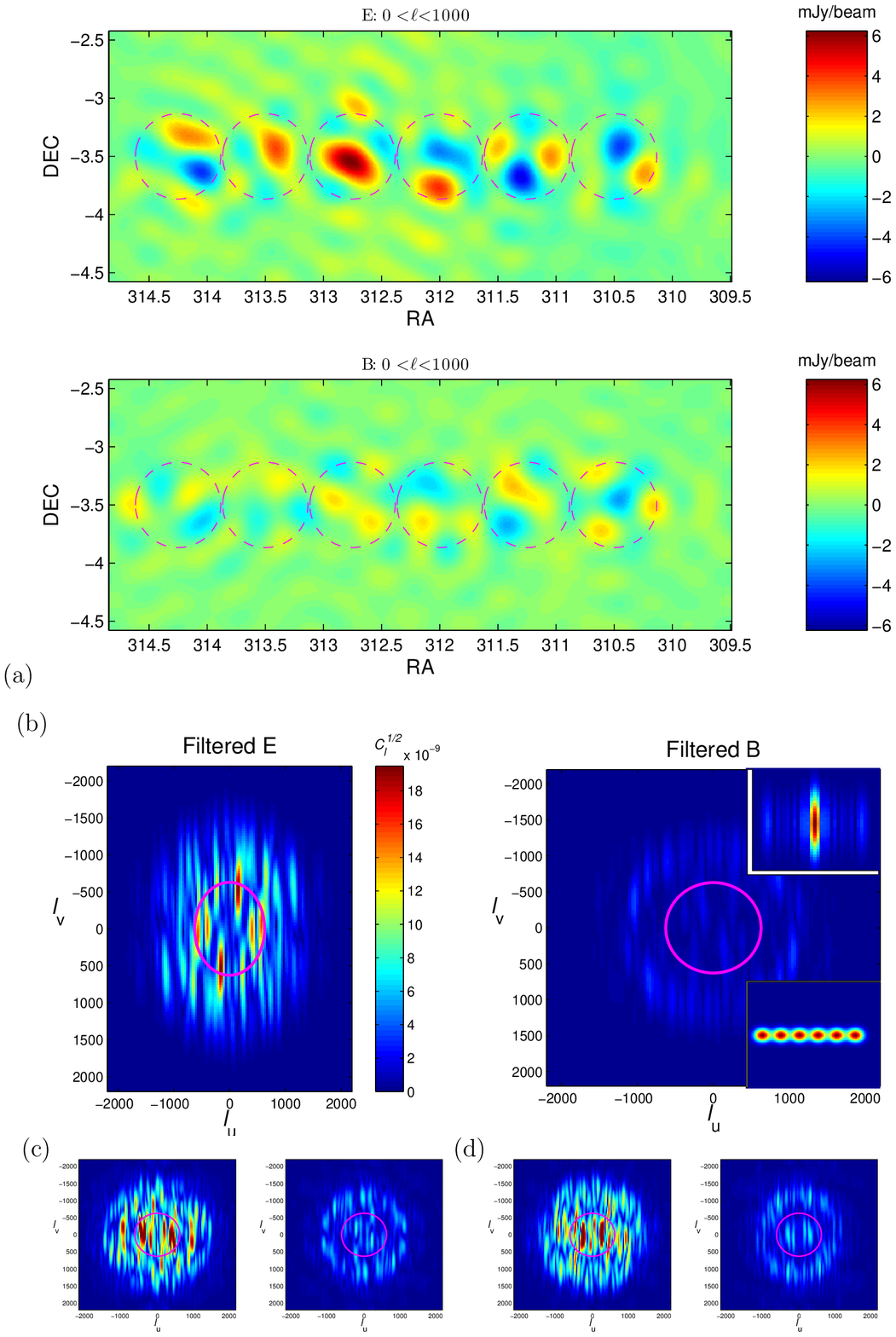}
\caption{ (a) Raw images of the (approximate) E and B signals from the
$20^h$ deep strip for $\ell < 1000$. The circles show the {\it fwhm}
of each of the six CBI pointings in the strip. A filter has been
applied to remove the mean-mode (ground) signal, which would otherwise
dominate the maps. No other filters are needed: the raw $20^h$ data
are dominated by the CMB signal.  For the data plotted here,
simulations of pure-E and pure-B signals show that the mixing between
E and B from sky coverage is $\sim$5\% in variance.  The E map, which
is predominantly signal, has a variance 2.8 times larger than that in
the B map (which is consistent with the noise).  Because Fourier
transforms preserve power, the E/B variance ratio is the same in
$\ell$ space.  (b) The modulus of the optimal (Wiener-filtered) signal
maps seen in the gridded $(\ell_u, \ell_v)$-plane using our signal
reconstruction process, where $(\ell_u, \ell_v) =2\pi (u,v))$.  The
maps have been normalized to $\sqrt{{\rm{C}}_\ell}$, equivalent to plotting
$a_{\ell,m}$'s for a full-sky map.  The
radius of the magenta circles is $\ell=630$, the peak of the CBI's
sensitivity.  The EE to BB power ratio in the Wiener-filtered,
reconstructed maps is 14.8.  The upper inset in the B map is the
average mosaic smoothing kernel ${\cal B}$ we chose to smooth the
reconstructed maps. The strong anisotropy in $\ell_u, \ell_v$ reflects
the geometry of the $20^h$ strip.  The lower inset is the sky plane
representation of $\cal{B}$.  (c,d) show two sample maps of the
fluctuations $\delta s = s - \langle s \vert \Delta \rangle $ on the same scale as (b).}

\label{fig:raw_EB_map}
\end{figure*}

\begin{figure*}[!t]
\centering

\includegraphics[height=0.8\textwidth, angle=0]{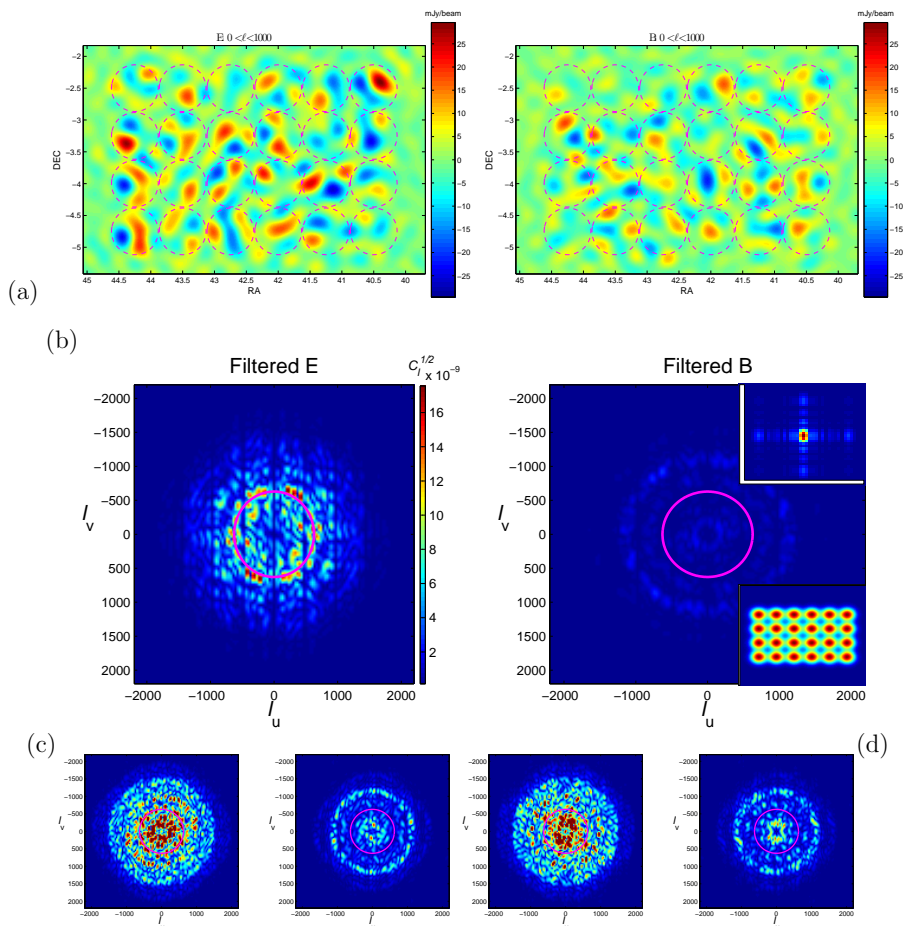}
\caption{ (a) Raw maps of the (approximate) E and B signals from the
$02^h$ mosaic field for $\ell < 1000$. The two
northern strips in this field have substantially less data than the
four southernmost strips; to keep the noisy northern strips from
visually dominating the map, they have not been included in the maps.
A filter
has been applied to remove the mean-mode (ground) signal, which would
otherwise dominate the maps.  The variance in the E map is 1.71 times
larger than the variance in the B map (which is consistent with the
noise).  The circles show the {\it fwhm} of each of the 24 CBI
pointings used in the maps (all 36 mosaic pointings are used in producing the spectrum).
(b) The modulus of the optimal (Wiener-filtered)
signal maps seen in the gridded $(\ell_u, \ell_v)$ plane
using our signal reconstruction process, where $(\ell_u, \ell_v)  =2\pi (u,v))$.  The
maps have been normalized to $\sqrt{{\rm C}_\ell}$, equivalent to plotting
$a_{\ell,m}$'s for a full-sky map. 
The radius of the magenta circles is $\ell=630$, the peak of the CBI's sensitivity.
The striping apparent in the E image is due to the
ground filter removing Fourier modes equal to the separation of the
mosaic fields in RA.  The EE to BB power ratio is 24. The ring at
$\ell \sim 1000$ in the B image is due to the $\sim 1.3\sigma$
bandpower value in the fifth BB bin which enters into the filter.  
The upper inset in the B map is the
average mosaic smoothing kernel ${\cal B}$ we chose to smooth the
reconstructed maps. The strong anisotropy in $\ell_u, \ell_v$ in
Fig.~\ref{fig:raw_EB_map} is not there in this square map.  The
compactness of ${\cal B}$ relative to that of
Fig.~\ref{fig:raw_EB_map} demonstrates the improved $\ell$-space resolution
of a mosaic, at a cost of signal to noise ratio.  The cross pattern
extending from the central peak of ${\cal B}$ arises from the Fourier
transform of the mosaic pattern on the sky.  The lower inset is the
sky plane representation of $\cal{B}$.  (c,d) show two sample
realizations of the fluctuations $\delta s = s - \langle s \vert
\Delta \rangle $ on the same scale as (b).}

\label{fig:raw_EB_map_square}
\end{figure*}

The gridded estimators allow us efficiently to reconstruct the polarization of
the CMB.  We use the fact that E and B are real field components that
completely describe the polarization field to present 
E and B sky images in
Figs.~\ref{fig:raw_EB_map}(a),~\ref{fig:raw_EB_map_square}(a).  
These images are
created by Fourier transforming the ground-filtered $\ell$-space 
 estimators $\Delta_{{\rm E}Q}$ and $\Delta_{{\rm B}Q}$.  These
$\Delta_{{\rm E}Q}$ and $\Delta_{{\rm B}Q}$ in 
turn are convolved representations of the true $\ell$-space $\tilde{\rm E}$ and
$\tilde{\rm B}$ which are related to the linear polarization Stokes parameters Q
and U \citep{Kovac:2002fg}.  These novel E and B images and $\ell$-space maps
differ from the traditional polarization ``headless'' vector plots based on decomposing the
images in Stokes Q and U into E-like and B-like components, as they
are direct representations of the E and B fields and their transforms.
Note that due to the non-local transformation relating E and B to the Q
and U fields on the sky, \eg \cite{Lewis02prd,Bunn:2002df}, the E and B images do not
show polarized ``objects'' but coherences in the polarization field.

The dominant ground contamination has been removed from the estimators by
forming $C_N(C_N+C_P)^{-1} \Delta$ in the large $C_P$ limit.  Here $C_N$
serves to regularize $C_P$.  Other regularizers, such as $C_N + C_T$, give
virtually identical results.
The images generated from the estimators include the effects
of the observing strategy and of the gridding process, which include the
mosaic pattern, $uv$ coverage, the primary beam, and noise-weighting, similar to a standard interferometric ``dirty map''.
Hence the raw images are not faithful reconstructions of true intensity 
and polarization.  
Furthermore, the Fourier
transform of the regular mosaic pattern introduces ``sidelobes'' in the
$\ell$-space map made from the raw estimators.  Therefore, filtering
and deconvolution can be beneficial to our maps.

\begin{figure*}[!t]
 \centering
\includegraphics[height=0.60\textwidth, angle=0]{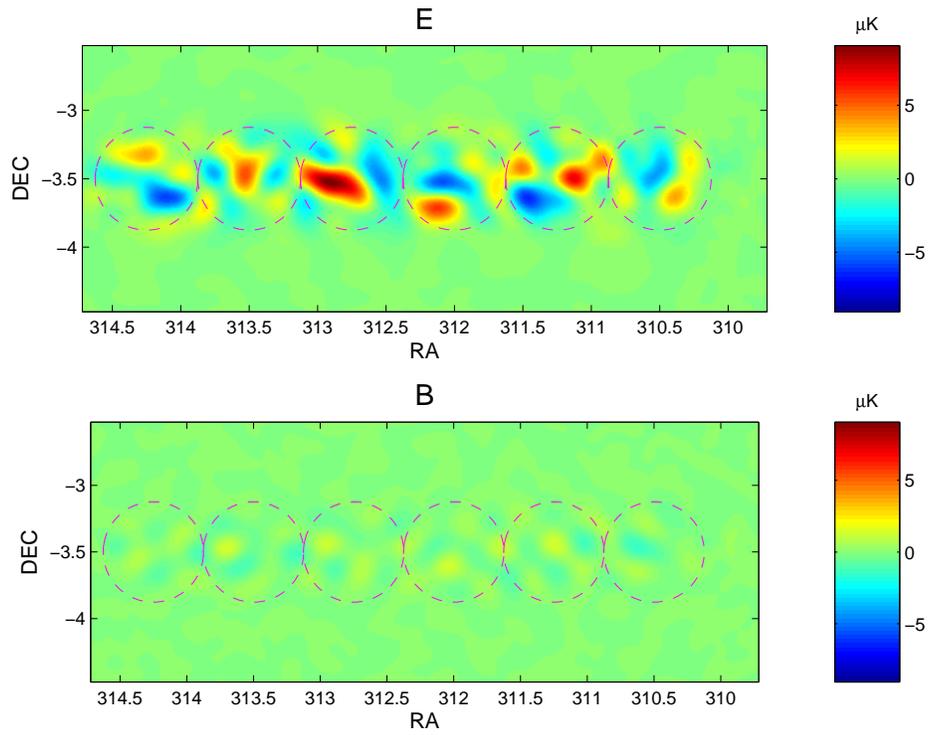}

\caption{ The E and B reconstructed signal sky-plane images of the
  $20^h$ deep strip, the transform of the $\ell$-space images in 
  Fig.~\ref{fig:raw_EB_map}(b).  These should be compared with the
  raw images in Fig.~\ref{fig:raw_EB_map}(a).  }
\label{fig:deconvolved_20hr_real}
\end{figure*}

Figs.~\ref{fig:raw_EB_map},\ref{fig:raw_EB_map_square}(b) show
optimal Wiener-filter $\ell$-space maps. These are smoothed mean
signals given the observations,
\begin{eqnarray} 
&&\langle s \vert \Delta \rangle = {\cal B} {\rm R}^{\dagger -1} C_T
C_{tot}^{-1} \Delta \, , \ C_{tot} \equiv C_N +C_P + C_T\, . \label{eq:sbar}
\end{eqnarray} 
Here the matrix ${\rm R}$ takes signal-space into data-space, with
$\Delta = {\rm R} s + n$, where $s$ is the signal, ${\cal B}$
is a smoothing kernel, and $n$ is the map-noise, including
ground and source projection terms.  One is free to choose the basis
in which to describe the true sky signals $s_i$, \eg as Stokes
parameters on the sky relative to a fixed sky-basis or as a set of
$a_{i\ell m}$ coefficients or (in the small angle limit) as the set of
related Fourier transform coefficients.  For the CBI, and
interferometers in general, it is natural to express the $s_i$ in the
$\ell$-basis, as in
Figs.~\ref{fig:raw_EB_map},\ref{fig:raw_EB_map_square}(b), but we also
show the sky plane version in Fig.~\ref{fig:deconvolved_20hr_real}.
The $C_T^{X} = \sum q_{Xb} C_{Tb}^{X} $ used in the signal
reconstructions are derived from the data using the measured
bandpowers $q_{Xb}$ (set to zero if negative) and the top-hat band
matrices $C_{Tb}^{X} $, and hence are ``theory-blind''.

Typically the signal-space dimension would be larger than the
data-space dimension so ${\rm R}$ would not be invertible.  By modeling
the $(\ell_u, \ell_v)$-plane only at points where we have estimators, the two
dimensions are equal, ${\rm R}$ becomes square, and is, in principle,
invertible. In practice ${\rm R}$ has an enormous condition number, so
we remove poorly measured modes with eigenvalues $< 10^{-4}$ of the
largest eigenvalue of ${\rm R}$ in constructing the ${\rm R}^{-1}$
operator. This makes the reconstruction better conditioned and the
results are insensitive to variations in the eigenvalue cut.  The
remaining ``noise'' in ${\rm R}^{-1}$ is controlled by reconvolution
with a smooth regularizer ${\cal B}$.

We have freedom in the choice of the smoothing kernel ${\cal B}$. The
one we choose for the images is a natural one associated with the
map. Letting the pixel be denoted by the vector ${\bf Q}
=(\ell_u,\ell_v)$, the matrix ${\rm R}$ has components {${\rm
R}${\boldmath{$_{Q, Q+\delta \ell}$}}}, where the vector
{\boldmath{$\delta \ell$}} goes over the region of $\ell$-space that
contributes to the pixel in question. ${\rm R}$ tends to be only roughly
independent of ${\bf Q}$.  We take our smoothing kernel {${\cal
B}${\boldmath{$_{\delta \ell}$}}} to be the average of {${\rm
R}${\boldmath{$_{Q, Q+\delta \ell}$}}}, over all pixels ${\bf Q}$.  As
can be seen in the insets in
Figs.~\ref{fig:raw_EB_map},\ref{fig:raw_EB_map_square}(b), this is
very nearly the Fourier 
transform of the mosaic pattern on the sky, as is required to make the
sky plane images reflect the area actually observed.  For the strip
{${\cal B}${\boldmath{$_{\delta \ell}$}}} is quite asymmetric, as
shown in Fig~\ref{fig:raw_EB_map}.  For the other three square CBI
polarization fields, the {${\cal B}${\boldmath{$_{\delta \ell}$}}} is
nearly symmetric, as shown in Fig~\ref{fig:raw_EB_map_square}.  There
are low-level "sidelobes" in the $\ell_u$ and $\ell_v$ directions due
to the mosaic spacing of $45^\prime$.
Figure~\ref{fig:deconvolved_20hr_real} shows the sky plane (as opposed
to $\ell$-space) representation of the reconstructed signal maps for
the deep strip.  The agreement between the raw E image in
Fig.~\ref{fig:raw_EB_map}(a) and the reconstructed E image in
Fig.~\ref{fig:deconvolved_20hr_real} is quite good.

To assess how well the mean field $\langle s \vert \Delta \rangle $
describes the actual distribution of signal on the sky, it is
important to see how large the fluctuations $\delta s \equiv s-
\langle s \vert \Delta \rangle $ are about it. For Gaussian signals
the statistics of the (smoothed) $\delta s$ are fully described by the
constrained correlation matrix
\begin{eqnarray} 
&&\langle \delta s \delta s^\dagger \vert \Delta\rangle = {\cal
  B} (w_S^{-1} -w_S^{-1} {\rm R}^\dagger C_{tot}^{-1}
  {\rm R}w_S^{-1}) {\cal B}^\dagger \label{eq:deltas} \\
&& = {\cal B} (w_S +  {\rm R}^\dagger (C_N +C_P)^{-1}
  {\rm R})^{-1} {\cal B}^\dagger\, . \nonumber  
\end{eqnarray} 
Here $w_S^{-1}$ is the correlation matrix of $s$, the unconstrained
signal in signal-space.  The signal variance in data-space is therefore
$C_T = {\rm R} w_S^{-1} {\rm R}^\dagger$. In terms of $w_S$, the
unsmoothed mean field is $\langle s \vert \Delta \rangle = w_S^{-1}
{\rm R}^\dagger C_{tot}^{-1} \Delta$. These equations apply to a
general ${\rm R}$, \eg when the signal-space and data-space dimensions
are unequal. If ${\rm R}$ is an invertible square matrix, then the
smoothed mean field is equation(\ref{eq:sbar}) and smoothed
realizations of the fluctuations are of form
\begin{eqnarray} 
&& \delta s = \langle \delta s \delta s^\dagger \vert
\Delta\rangle^{\half} g \, , \label{eq:deltasrealize} \\ 
&& \langle \delta s \delta s^\dagger \vert \Delta\rangle = 
{\cal B}\ {\rm R}^{-1} (C_T - C_T C_{tot}^{-1}C_T) {\rm R}^{\dagger
-1}{\cal B}^\dagger\, .
\label{eq:deltascorr} 
\end{eqnarray} 
Here $g_{iQ}$ are independent Gaussian random variables with unit
variance and $\langle \delta s \delta s^\dagger \vert
\Delta\rangle^{\half}$ is the matrix square root.  Equation
(\ref{eq:deltascorr}) shows the fluctuations go to zero in modes in
which the generalized noise is small, but approach pure signal
realizations in modes in which it is
high. Figs.~\ref{fig:raw_EB_map},~\ref{fig:raw_EB_map_square}(c,d)
show a few examples using the $C_T$ derived for the four CBI
fields. These illustrate that the reconstructed signal for the deep
strip is better determined than for the mosaic fields.

We have also used a modified version of the CLEAN deconvolution
algorithm \citep{clean74} to do the effective inversion of ${\rm R}$:
we find the largest signal among the $uv$ estimators $\Delta_{iQ}$; we
place a $\delta$-function in $\ell$-space there that zeroes out the
$\Delta_{iQ}$; we subtract from each estimator its response to that
signal; we then repeat, ending only when $\sim 10^4$ components have
been found. This leads to an error in the residual less than $10^{-5}$
of the power in the original. This method has several nice features:
since each estimator has compact support in the $uv$ plane, the
process is quite stable because the addition of a model component only
affects a few estimators; since the model is in the same space as the
data, no (time-consuming) Fourier transforms or 
(expensive) decomposition of ${\rm R}$ are needed.  As in the standard
interferometer imaging application of the CLEAN algorithm, a smooth
restoring convolution kernel ${\cal B}$ is required to turn the set of
$\delta$-functions in $\ell$-space resulting from the CLEANing into a smooth
transformable map, otherwise the image would have artifacts.

\begin{figure*}[!t]
 \centering
\includegraphics[height=0.9\textwidth, angle=0]{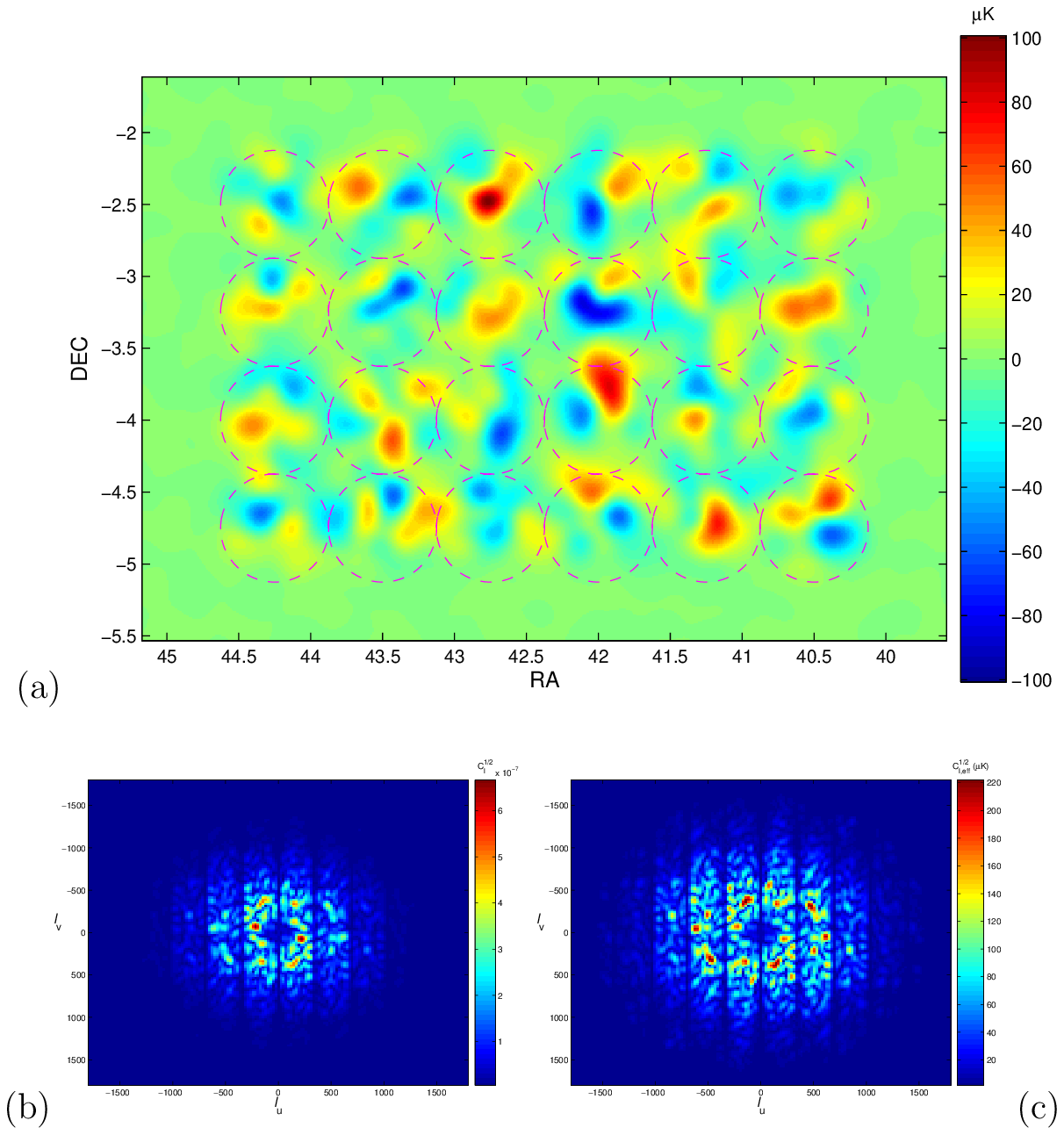}
\caption{ (a) The reconstructed sky image for the Wiener-filtered
  $02^h$ mosaic in total intensity.  Both ground and sources (which
  would otherwise dominate the map) have been removed as part of the
  filter.  This image has been produced using the $uv$-plane CLEAN
  procedure described in \S~\ref{sec:maps}.  The circles show the {\it
  fwhm} of each of the 24 CBI pointings used in
  Fig.~\ref{fig:raw_EB_map_square}.  .(b) The modulus of the
  optimal (Wiener-filtered) map of (a) in the gridded
  ($\ell_u,\ell_v$) plane.  As in Fig.~\ref{fig:raw_EB_map_square}(b), the vertical
  striping is due to the ground filter.  (c) A rescaling of (b) that
  brings out the structure in the $\ell_u,\ell_v$-plane by multiplying each
  pixel by $\ell$.  This is equivalent to using
  $\ell(\ell+1)\rm{C}_\ell/2\pi$ instead of $\rm{C}_\ell$ to plot the
  angular power spectrum.  In this representation, one can see hints
  of multiple rings of higher signal that correspond to the Doppler
  peaks in the power spectrum. }
\label{fig:deconvolved_02hr_t}
\end{figure*}

We find that the results using eigenvalue cuts or this CLEAN method
give very similar maps.  The sky plane and $\ell$-space total
intensity images of the $02^h$ CBI mosaic which we display in
Fig~\ref{fig:deconvolved_02hr_t} were constructed using this CLEAN
algorithm. As expected given the relatively small errors on the T
bandpowers, all four fields show strong T detections.

\section{Parameterized Polarization Phenomenology}
\label{sec:params}

\subsection{TT, EE, TE and BB Bandpower Data}
\label{sec:bandpowdat}

In our parameter determinations we consider five combinations of
bandpower data: (1) CBI TT+EE+TE bandpowers obtained from the analysis
described here of the 2002--2005 data with a bin width $\Delta \ell
\approx 75$, more fine-grained than those of Table~\ref{tab:7bandpows}
(available as an on-line supplement); (2) CBI $\ell=600$ to
$\ell=1960$ TT bands from the combined mosaic and deep field analysis
of \cite{Readhead04}; (3) TT and TE WMAP1 bandpowers from the first
year WMAP data \footnote{None of the conclusions drawn here are
affected by using the WMAP 3-year power spectrum (released after
submission of these results).}  \citep{Bennett03}, adopting the
likelihood mapping procedure described in \cite{Verde03}; (4) DASI TT
\citep{Kovac:2002fg} and 3-year EE+TE \citep{Leitch04} results; (5)
the recent Boomerang B03 TT+EE+TE results
\citep{Masi:2005,Jones:2005,Montroy:2005,Piacentini:2005}, with $\ell
< 300$ TT bandpowers excluded because of overlap with WMAP (although
this has no quantitative impact).  We also omit CBI TT results for
$\ell < 360$ both because of the overlap with WMAP, and because the
(very limited) sensitivity there is coming from the sidelobes of the
primary beam which are extremely difficult to measure accurately.  The
Markov chain Monte Carlo (MCMC) package \COSMOMC
\footnote{\url{http://cosmologist.info/cosmomc} } 
\citep{Lewis02}, modified to 
include polarization spectra, the cross correlation between TT and EE
spectra, and isocurvature modes, is used to calculate posterior
probability distributions (including priors) for cosmic parameters.

\subsection{The Basic  Flat Tilted Adiabatic  $\Lambda$CDM Model}
\label{sec:ad_params} 

The simplest inflationary paradigm is characterized by six basic
parameters: $\omega_b\equiv\Omega_bh^2$, the physical density of
baryons; $\omega_c\equiv\Omega_ch^2$, the physical density of cold
dark matter; $\theta\equiv100\ell_s^{-1}$, parameterizing the angular
scale $\ell_s^{-1}$ associated with sound crossing at decoupling,
which defines the overall position of the peak--dip pattern; $\ln(
10^{10} A_s)$, the logarithm of the overall scalar curvature
perturbation amplitude $A_s$, the scalar curvature power spectrum
${\cal P}_{{\rm s}} (k)$ evaluated at the pivot point $k_n=0.05 \Mpc^{-1}$;
$n_s$, the spectral index of the scalar perturbations, defined by
${\cal P}_{{\rm s}} (k) \propto k^{n_s -1}$ ; and $\tau$, the Thomson
scattering depth to decoupling. We do not consider gravitational-wave
induced components.

Table~\ref{tab:base} shows the broad priors we have chosen for the
basic parameter ranges so that they have little influence on our
results. We also impose a weak-$h$ prior on the Hubble constant $H_0
=100 h \,\rm km\,s^{-1}\,Mpc^{-1}$: $0.4 < h < 1$.  For the flat
$\Omega_{\rm tot}=1$ models considered here this weak-$h$ prior
has little influence on the results, although some extreme models with
high Thomson depth are excluded. The strongest prior is the flat
restriction, expected in most inflation models. Some parameters change
significantly when the flat prior is relaxed
\citep{Bond03,Readhead04}.

We highlight distributions for two parameters, the pattern-shifting
$\theta/\theta_0$ and $q_s = A_se^{-2\tau}/A_{s0}e^{-2\tau_{0}}$ which
determines the overall ${\cal C}^{X}_\ell$ amplitude. We normalize
relative to $\theta_0=1.0442$ and $A_{s0}e^{-2\tau_0} = 18.0 \times
10^{-10}$, the best-fit values for WMAP1+CBI+DASI+B03 TT+EE+TE
(Table~\ref{tab:base}). The near-degeneracy between $A_s$ and $\tau$
is only weakly broken at very low $\ell$ where reionization has some influence,
and at higher $\ell$ through (nonlinear) secondary phenomena such as
weak lensing or the Sunyaev-Zeldovich effect. However, the $q_s$
combination is reasonably well determined, although there are
correlations with other parameters, \eg $\omega_b, \omega_c$ and
$n_s$, especially with polarization data only.

The relative positions of the peaks in the TT and EE spectra are
``phase-locked'' by the physics of the acoustic oscillations at photon
decoupling, with the multipole of the $j^{th}$ TT peak $\propto j
\theta^{-1}$, and the multipole of the $j^{th}$ EE peak $\propto
(j+1/2)\theta^{-1}$. The TE cross-spectrum has double the number of
peaks (Fig.~\ref{fig:7BinSpec}). For WMAP1+CBI+DASI+B03 TT+EE+TE, we
find $\theta/\theta_0= 1.001 \pm 0.0041$ and $q_s= 0.996 \pm
0.037$. For WMAP1+CBI $\theta/\theta_0= 0.999 \pm 0.005$, to be
contrasted with the $1.000 \pm 0.005$ obtained in \cite{paper8} for
this $\theta_0$.  These results for the mean values and standard
deviations are also very similar to those obtained with other CMB data
combinations, \eg \cite{Bond03,MacTavish:2005}.

\begin{figure*}[!t]
\centering
\includegraphics[width=0.7\textwidth]{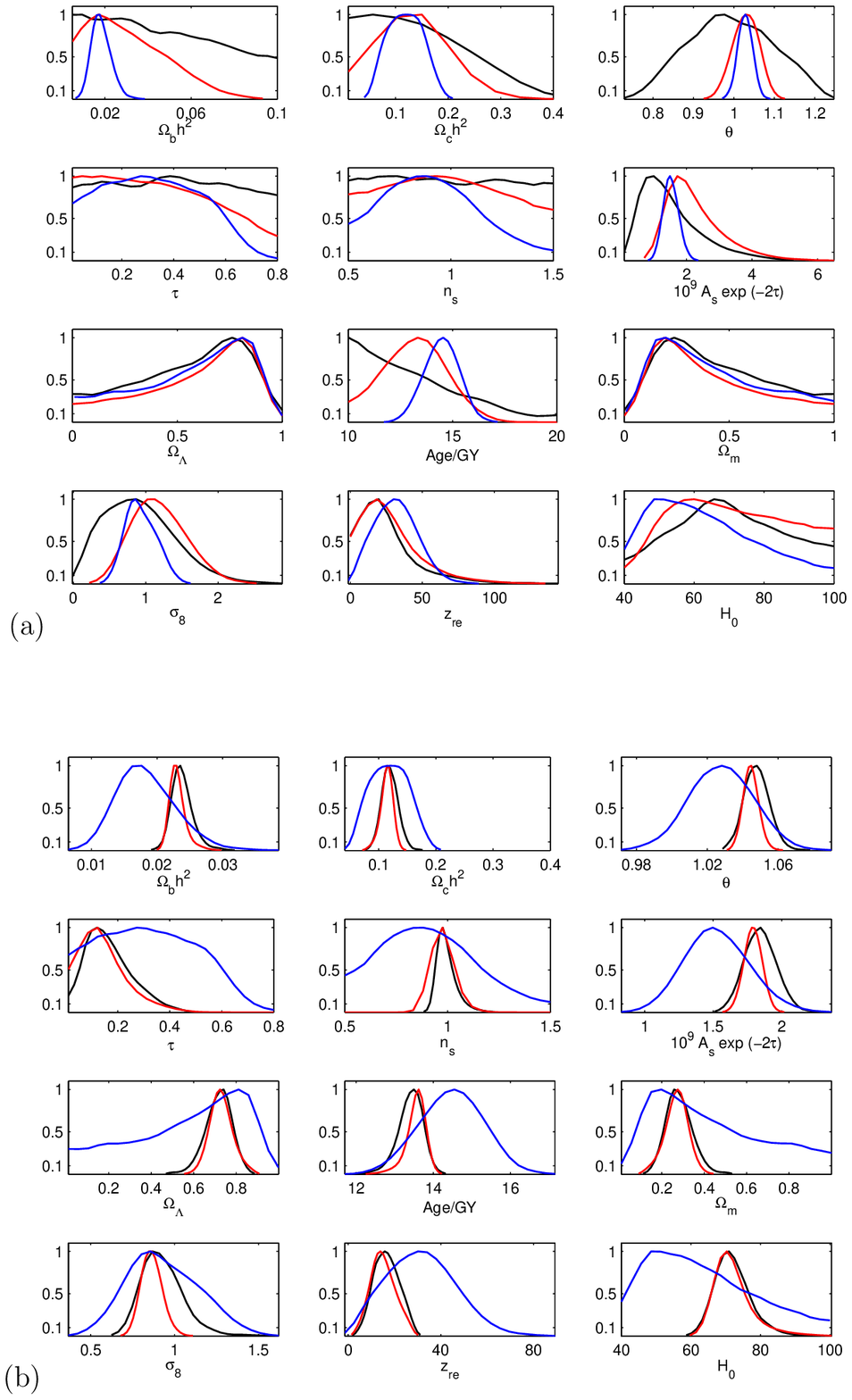} 
\caption{One-dimensional likelihoods for the cosmological parameter
indicated, marginalized over the other parameters. (a) shows CBI EE
(red), CBI+DASI+B03 EE+TE (blue). The black line is the
prior, calculated using a parameter run with CBI BB only. (b) shows
WMAP1 only TT+TE (black), WMAP1 TT+TE + CBI+DASI+B03 TT+EE+TE (red) and
CBI+DASI+B03 EE+TE 
(blue).} 
\label{fig:cbipol_dasi_b03_params}
\end{figure*}

For WMAP1+CBI+DASI+B03 TT+EE+TE the other basic cosmic parameters,
after marginalization, have distributions shown in
Fig.~\ref{fig:cbipol_dasi_b03_params} with median values and 1$\sigma$
errors given in Table~\ref{tab:base}.  The best-fit parameters for the
fiducial model are also given there. Using WMAP1 TT+TE and
CBI+DASI+B03 TT gives very similar results: the inclusion of the
current high $\ell$ polarization data has little impact
on parameter values for this limited set.

For CBI+DASI+B03 EE+TE, we get $\theta/\theta_0$ = $0.986 \pm 0.017$
and $q_s= 0.86 \pm 0.14$, in good agreement with the TT result. (A few
other parameters are also moderately well constrained, but most are
not, as shown in Table~\ref{tab:base}.)  Our $\theta/\theta_0$ result
is not affected if we relax the flat prior ($0.988 \pm 0.018$). For
CBI EE $\theta/\theta_0$ is $0.986 \pm 0.031$ and $q_{s}$ is $1.27
\pm 0.48$.  For CBI EE+TE $\theta/\theta_0$ is $0.95 \pm 0.027$ and
for DASI+B03 EE+TE it is $1.03 \pm 0.030$. For CBI BB we obtain
$\theta/\theta_0 = 0.95 \pm 0.10$. This should be interpreted as
essentially the limit that we would get from the prior probabilities
alone.  This shows our $\theta/\theta_0$ results are data-driven
rather than prior-driven.

\begin{table*}
\centering
\space2
\caption{Cosmic parameter values for the flat tilted adiabatic
  $\Lambda$CDM model}
\label{tab:base} 
\space
\scriptsize
\begin{tabular}{|c||c|c|c|c|}
\hline\hline
& & & &  \\
  & prior & WMAP1+CBI+DASI+B03  & WMAP1+CBI+DASI+B03  & CBI+DASI+B03  \\					
 & range & TT+EE+TE (best-fit) & TT+EE+TE & EE+TE \\
& & & &   \\				
\hline
 & & & & \\
$\theta / \theta_0$    & 0.5 to 10 & 1          & $1.001 \pm 0.0042$   & $0.987 \pm 0.017$ \\
$       \Omega_b h^2 $ & 0.005 to 0.1 &  0.0226 & $0.0232 \pm 0.0013$ & $0.018 \pm 0.005$	\\
$       \Omega_c h^2 $ & 0.01 to 0.99 &  0.117  & $0.114 \pm 0.011$    & $0.119 \pm 0.034$	\\
$           \tau     $ & 0.01 to 0.8  &  0.105  & $0.149 \pm 0.086$    & $0.33 \pm 0.18$	\\		    	
$         n_{\rm s}  $ & 0.5 - 1.5    &  0.960  & $0.978 \pm 0.039$    & $0.92 \pm 0.23$ 	\\
$\ln[10^{10} A_{\rm s}]$ & 2.7 to 4.0 &  3.09   & $3.18 \pm 0.16$      & $3.37 \pm 0.35$	\\
& & & &   \\				
\hline
& & & &   \\		
$q_s = A_se^{-2\tau}/A_{s0}e^{-2\tau_0}$ & -         &  1    & $0.992 \pm 0.037$  & $0.86 \pm 0.14$ \\
$ \Omega_\Lambda                       $ & -  	     & 0.714 & $0.733 \pm 0.054$  & $0.58 \pm 0.25$ \\
$     {\rm Age(Gyr)}                   $ & -         & 13.6  & $13.5 \pm 0.26$    & $14.4 \pm 0.80$ 	\\
$       \Omega_m                       $ & -         & 0.286 & $ 0.267\pm 0.054$ & $0.42 \pm 0.25$  \\
$       \sigma_8                       $ & -         & 0.83 & $ 0.848 \pm 0.063$ & $0.94 \pm 0.21$ \\
$         z_{re}                       $ & -         & 12.5  & $ 15.1 \pm 5.3$    & $32 \pm 15$ \\
$         H_0                          $ & 40 to 100 & 70.0  & $72.6 \pm 5.6$     & $64 \pm 15$ \\
\hline\hline
\end{tabular}
\small
{\begin{flushleft}
{The first group shows the six independent
(fitted) parameters, the second group shows parameters derived from
them.  Mean values and standard deviations are given for TT+EE+TE data
in column 4 and for EE+TE in column 5.
The ranges for the uniform weak priors we imposed for the MCMC runs
are given in column 2.  The best-fit model parameters defining our
``fiducial model'' are shown in column 3. For this model
$\theta_0=1.0437$ and $A_{s0}e^{-2\tau_0} = 17.9 \times
10^{-10}$. Here These are slightly different than the parameters
defining the WMAP team's best-fit \citep{Spergel03} using WMAP1 TT+TE
+ ACBAR TT + an earlier version of the CBI TT data \citep{Pearson03}
and different priors: $\Omega_b h^2=0.0224$, $\Omega_c h^2=0.111$,
$n_{\rm s}(k=0.05) = 0.958$, $\tau=0.11$, $H_0= 72$. This was the
fiducial model used in \cite{paper8}. Fig.~\ref{fig:7BinSpec} shows
that the two are very similar visually.}
\end{flushleft}}
\end{table*}

\subsection{A Peak/Dip Pattern Test}
\label{sec:ad_pattern} 

Single-band or broad-band results using theoretically motivated ${\cal
C}^X_{\ell b}$ shapes can also be produced by our pipeline. These
allow complete mapping of the full likelihood surfaces without using
the compressed bandpowers. The one-band model ${\cal C}_\ell^{X} = q_s
{\cal C}_{\ell}^{X({\rm s})}$, with ${\cal C}_{\ell}^{X({\rm s})}$ the fiducial
adiabatic model, yields for CBI EE $q_s=1.02\pm 0.14$ (68\%) and a
$11.7\,\sigma$ detection relative to $q_s=0$; CBI EE+TE gives a $12.4\,\sigma$
detection.  These can be compared with the 
$6.3\,\sigma$ DASI EE detection reported in \cite{Leitch04} and the
$8.9\,\sigma$ CBI EE detection reported in \cite{paper8} (with no
polarization point sources projected out, $7.0\,\sigma$ with 20\%
removal).  Alone, the new CBI TE data give $q_s=1.02\pm 0.24$ and a
$4.2\,\sigma$ detection relative to $q_s=0$.  The CBI TT data yield 
$q_s=1.12\pm 0.05$ which is a $95\,\sigma$ significance detection versus
$q_s=0$.

\begin{figure*}[!t]
\centering
\includegraphics[width=0.6\textwidth]{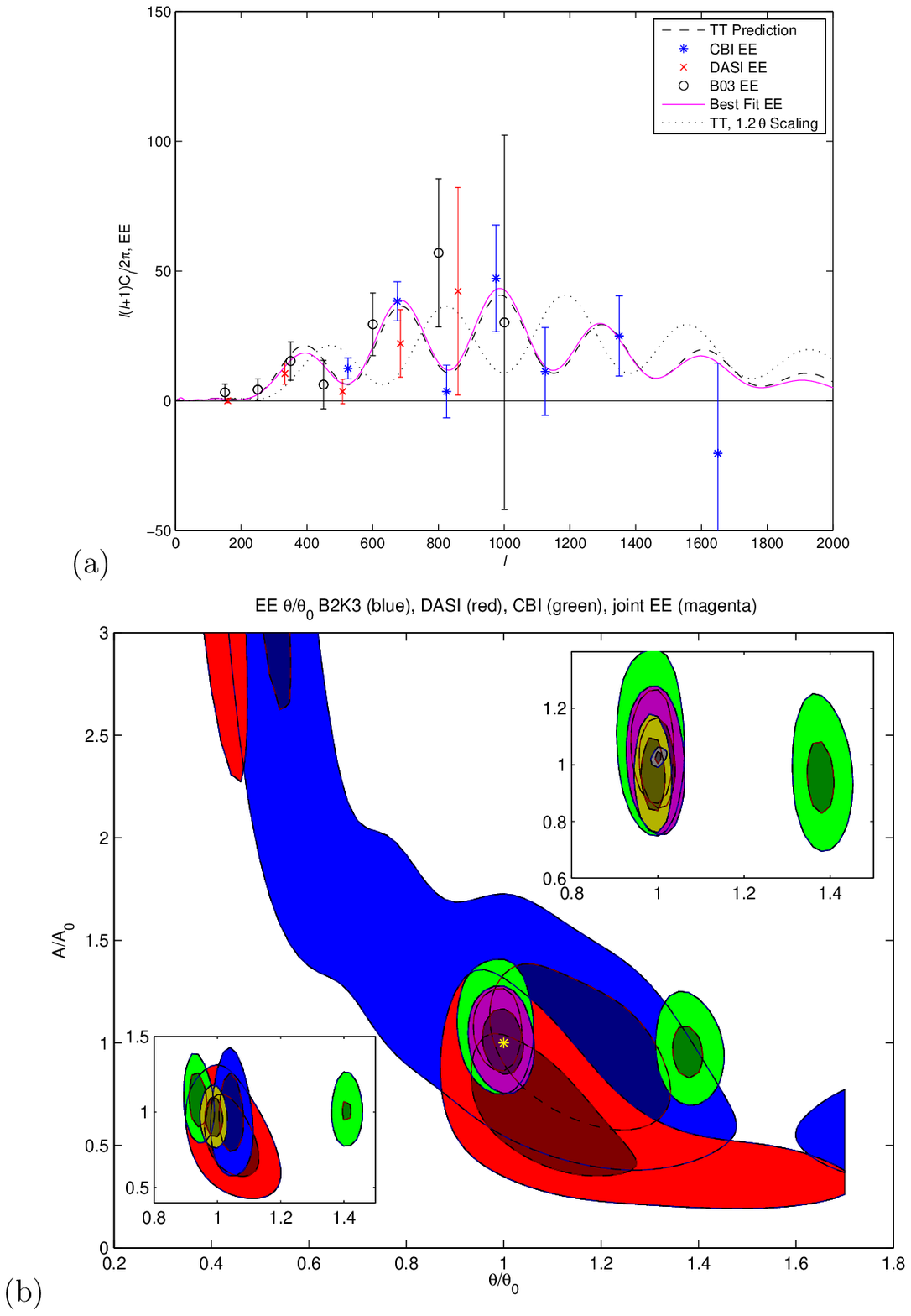} 
\caption{(a) shows the polarization data from CBI, DASI, and B03,
along with the fiducial EE prediction (black dashed), the same model
shifted by 20\% in $\theta$ (black dotted), and the best-fit
CBI+DASI+B03 EE cosmology (magenta).  The polarization data pick
out the same sound-crossing angular scale as the TT data do, with
$\theta/\theta_0=$ $0.986 \pm 0.017$ when marginalized over the other
5 cosmological parameters.  (b) shows 1 and 2 $\sigma$ contours
derived from CBI EE (green), DASI EE (red), B03 EE (blue) and
CBI+DASI+B03 EE (magenta) for the 2-parameter $q_s$ and
$\theta/\theta_{0}$ template model.  Marginalization yields
$\theta/\theta_0 = 0.993 \pm 0.027$, $q_s = 1.00 \pm 0.11$. The yellow
star marks the expected result from the fiducial model. The lower
inset shows the same with EE+TE instead of EE only. The upper inset
shows the sequence CBI EE (green), CBI+DASI+B03 EE (magenta),
CBI+DASI+B03 EE+TE (brown). Marginalization yields $\theta/\theta_0 =
0.988 \pm 0.018$, $q_s = 0.97 \pm 0.09$. Grey contours denote
CBI+B03 TT.}
\label{fig:theta_model_plot}
\end{figure*}

To further complement the MCMC determinations, we consider the
two-parameter template model\footnote{In \cite{paper8} we also
described a 2-parameter ``sliding comb'' test of the
phase-relationship between TT and EE. This involved an underlying
smooth ${\cal C}_{\ell}^{X({\rm s})}$ with a sinusoidal pattern
characterized by an angular phase shift $\phi$ designed to give the
fiducial model forecast for EE when $\phi =0$. The best-fit CBI EE
phase was $21^{\circ}\pm 40^\circ$ with amplitude $q_s=1.07 \pm 0.21$;
the new data give $13^{\circ}\pm 36^\circ$ and $1.05 \pm 0.13$.}
${\cal C}_\ell^{X} = q_s {\cal C}_{\ell(\theta_0 /\theta)}^{X({\rm
s})}$, evaluated on a grid in $(q_s,\theta/\theta_0)$. The other
cosmic parameters are fixed at the fiducial model values.  We restrict
$\theta /\theta_0$ to lie between $0.3$ and $1.7$, the range of our
grid.  Fig.~\ref{fig:theta_model_plot}(a) shows how the EE peak/dip
pattern shifts for the polarization.  The $\theta/ \theta_0$--$q_s$
likelihood contours in Fig.~\ref{fig:theta_model_plot}(b,c,d) show
that for each of the EE polarization datasets there is a multimodal
probability structure. For example, for CBI, apart from the $\theta/
\theta_0 \approx 1$ solution, there is another with the third
polarization peak shifted and scaled to fit the second peak of the
fiducial model. There is a strong probability minimum in between the
two. This multiple solution disappears when DASI and B03 are combined
with CBI, yielding the well-determined $\theta/\theta_0 = 0.988 \pm
0.018$, $q_s = 0.97 \pm 0.09$ for EE+TE. These are in good agreement
with the $0.986 \pm 0.017$ and $q_s= 0.86 \pm 0.14$ MCMC numbers
determined by marginalizing over the other five cosmic parameters. The
multimodal aspect is strongly suppressed in MCMC just because of the
extremely weak-$h$ prior we impose, but correlations among parameters
lead to the larger errors in $q_s$. For EE alone, the template grid
gives $\theta/\theta_0 = 0.993 \pm 0.027$, $q_s = 1.00 \pm 0.11$ and
the marginalized MCMC gives $\theta/\theta_0= 1.003 \pm 0.027$, $q_s=
1.07 \pm 0.30$.

Fig.~\ref{fig:theta_model_plot} shows the best-fit EE+TE power
spectrum. It looks remarkably like the TT fiducial model forecast.

\subsection{Adding a Subdominant Isocurvature Mode}
\label{sec:iso_params} 

Isocurvature modes which could lead to a measurable signal in the CMB
may arise in multiple scalar field models during inflation or they can
be generated after inflation has ended.  Necessary ingredients to
impact CMB and large scale structure (LSS) observations include:
association with a component of significant mass-energy, such as
baryons, cold dark matter, or, possibly, massive neutrinos (hot dark
matter); sufficiently large primordial fluctuations in the
entropy-per-baryon, the entropy-per-CDM-particle or the
entropy-per-neutrino.  A concrete CDM realization is the axion.  For
cosmic-defect-induced isocurvature perturbations, which could also
arise near the end of or after inflation, the mass-energy is in the
defects.

Isocurvature perturbations from quantum zero point fluctuations in
inflation have a well-defined pattern
\citep{1986MNRAS.218..103E,Bond:1987ub} in TT, EE and TE, with no BB
predicted (except through lensing). The peaks and dips are predicted
to be out of phase with those from adiabatic modes, as shown in
Fig.~\ref{fig:eeisocCL}.  (For defects it is difficult to get any
peaks and troughs at all.) Further, there is a large ``isocurvature
effect'' predicted at lower $\ell$ relative to that at high $\ell$
where the peaks are. 

A pure isocurvature mode does not fit the current TT data unless the
primordial isocurvature power spectrum ${\cal P}_{{\rm iso}} (k)$ is
designed to mimic the observed ${\cal C}^{{\rm TT}}_\ell$ pattern with its
own peak/dip structure and overall $k$-dependent blue tilt. Although
highly baroque in terms of inflation models, such radically broken
scale invariance is possible for isocurvature perturbations just as it
is for the adiabatic case. Polarization (and LSS) data help by
breaking such severe degeneracies with the cosmic parameters.  

A detailed analysis of a general set of isocurvature initial
conditions for four cosmological fluids using CMB and LSS data has
been undertaken in \cite{Moodley:2004}.  If one includes all allowed
isocurvature and adiabatic perturbations, and correlations between
them, the current CMB and LSS data still allow a substantial amount of
isocurvature perturbations.  However simpler and more
realistic models that only include an isocurvature perturbation in
one fluid are more strongly constrained. 

Here we assume Gaussian-distributed CDM isocurvature perturbations and
add two extra parameters beyond our base adiabatic six: two amplitude
ratios, $R_i \equiv {\cal P}_{{\rm iso}} (k_i)/{\cal P}_{{\rm s}} (k_i)$, at two
pivot wavenumbers $k_i$, one at small scale, $k_2=k_n =0.05 \Mpc^{-1}$
and one at large scale, $k_1=0.005 \Mpc^{-1}$. A (constant) primordial
spectral index defined by ${\cal P}_{{\rm iso}} (k) \propto k^{n_{iso}}$
follows: $n_{\rm iso}$ = $n_{\rm s} -1 + \ln (R_2/R_1)/ \ln
(k_2/k_1)$.

We find that for neither $R_i$ parameter is there evidence for an
isocurvature detection, in agreement with \cite{MacTavish:2005} who
used the same $R_1$--$R_2$ parameterization. We find for
WMAP1+CBI+DASI+B03 TT+EE+TE, 95\% confidence upper limits of $R_1$
$<0.26$ and $R_2 < 1.7$ on the higher wavenumber scales which \CBI\
probes. This translates into steeper $n_{\rm iso}$ values being more
allowed than the $n_{\rm iso} \approx 0 $ nearly scale invariant ones.
The CBI EE+TE data only limits $R_1 <18$ and $R_2 < 54$, whereas
CBI+DASI+B03 EE+TE gives $R_1 < 9$ and $R_2 < 30$.

Inflation models more naturally produce nearly scale invariant
isocurvature spectra, with $n_{\rm iso} \approx 0$, just as one often
gets $n_{\rm{s}} \approx 1$ for adiabatic perturbations. The tilts
from theory are also more likely to be red ($n_{iso} <0$) than blue
($n_{iso} > 0$).  However the data more strongly constrain red models
than blue.  

\subsection{Constraints on Interloper Isocurvature Peaks}
\label{sec:iso_pattern} 

\begin{figure*}[!t]
 \centering
\includegraphics[width=0.6\textwidth]{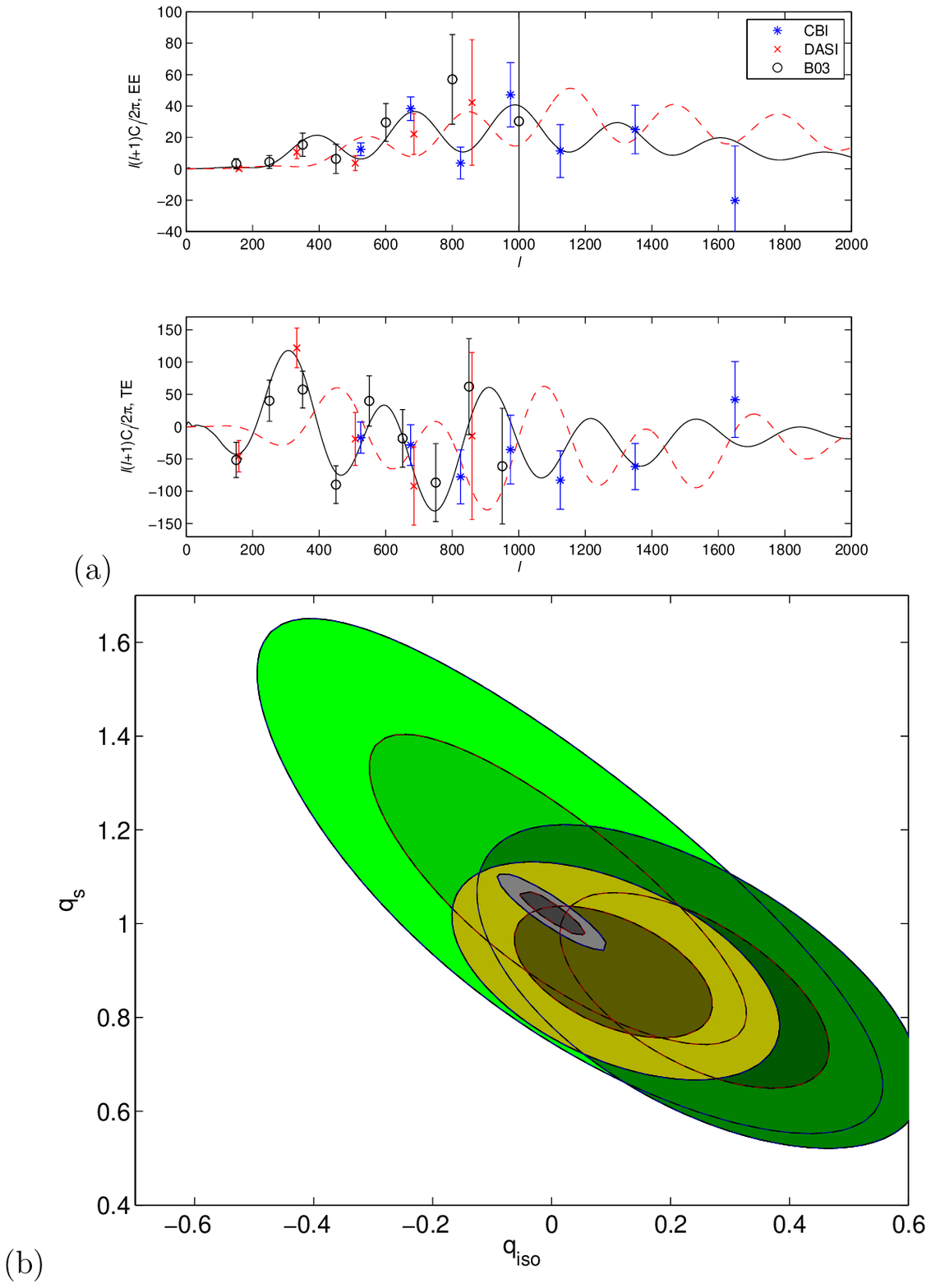}
\caption{(a) ${\cal C}^{{\rm EE}({\rm iso})}_{\ell}$ and ${\cal
  C}^{{\rm TE}({\rm iso})}_{\ell}$ power spectra for the $n_{iso}=3$
  white noise isocurvature model (red dashed) are compared with ${\cal
  C}^{{\rm EE}({\rm s})}_{\ell}$ and ${\cal C}^{{\rm TE} ({\rm
  s})}_{\ell}$ for the best-fit adiabatic fiducial model (black).
  These are the template spectra used for the 2-parameter
  $q_s$-$q_{iso}$ test. $q_{iso}$ is normalized to give the fraction
  of the expected EE power that is observed in the data over the range
  $400 \le \ell \le 1200$.  The CBI (blue asterisks), DASI (red x's)
  and B03 (black circles) EE and TE data are also shown.  (b) The
  likelihood surface for the 2-parameter $q_s$-$q_{iso}$ model, for
  the CBI EE (light green), CBI EE+TE (dark green), CBI+DASI+B03 EE+TE
  (brown), and CBI+B03 TT (grey) data.  The data strongly prefer the
  adiabatic over the isocurvature spectrum. Marginalization over the
  2D distributions yields: for CBI EE $q_s=1.05 \pm 0.22$,
  $q_{iso}=-0.01 \pm 0.21$; for CBI TE $q_s=0.81 \pm 0.24$,
  $q_{iso}=0.49 \pm 0.26$; and for CBI+DASI+B03 EE+TE data $q_s=0.90
  \pm 0.10$, $q_{iso}=0.10 \pm 0.11$.  
The polarization data are consistent with a single component adiabatic inflation model. }
\label{fig:eeisocCL}
\end{figure*}

To focus attention on the high $\ell$ polarization
results, we now fix $n_{\rm iso}$ to be the extremely blue 3, the
white noise `isocurvature' spectrum, with no spatial correlation.
Large angular scales in ${\cal C}^{X(iso)}_{\ell}$ are highly
suppressed and the isocurvature peaks and troughs emerge looking
somewhat like an $\ell$-shifted version of the adiabatic spectrum, as
shown in Fig.~\ref{fig:eeisocCL}. (The $n_{iso}=2$ case, which looks
even more like a shifted version of the fiducial model, gives similar
results to those given here.  See \cite{MacTavish:2005} for a
treatment of both $n_{iso}=2$ and $3$ cases.)  Although ${\cal
P}_{iso} (k) $ is so steep for such blue spectra that it must be
regulated by a cutoff at high $k >> k_2$, ${\cal C}^{X(iso)}_{\ell}$
has a larger natural damping scale so we do not need to add another
parameter.

The 2-parameter template model, ${\cal C}^{X}_{\ell} = q_s {\cal
C}^{X({\rm s})}_{\ell} + q_{iso}{\cal C}^{X({\rm iso})}_{\ell}$, 
therefore tests
at what level an interloper set of isocurvature peaks would be allowed
by the CMB data which, as we have seen in the $\theta/\theta_0$ test,
prefer the adiabatic peak positions. We normalize $q_{iso}$ so that
$q_{iso}=1$ corresponds to the same power in ${\cal C}^{{\rm EE}({\rm
iso})}_{\ell}$ as in ${\cal C}^{{\rm EE}({\rm s})}_{\ell}$ over a band
from $\ell = 400$ to $1200$. We find $q_{iso} \approx R_2
q_s/80$. Fig.~\ref{fig:eeisocCL}(b) shows a strong preference for the
pure adiabatic mode and no isocurvature detection, with $q_{iso}
=-0.01 \pm 0.21$ for CBI EE, $0.24 \pm 0.15$ for CBI EE+TE and $0.10
\pm 0.11$ for all of the polarization data.

We also let the full 7 cosmological parameters vary, using \COSMOMC\
to evaluate the probability distribution for $R_2$. This is a
different exercise than the 2-parameter case: to match the data, the
other parameters are adjusted by \COSMOMC\ to make the isocurvature
troughs and peaks interfere with the adiabatic peaks and troughs,
respectively, to mimic no interloping at all.  For CBI EE+TE we get
$R_2 < 76$ whereas for CBI+DASI+B03 EE+TE we get $R_2 < 44$. For
WMAP1+CBI+DASI+B03 TT+EE+TE, we get $R_2 < 3.0$.  We note that ${\cal
C}_B^{{\rm TT}({\rm iso})} / {\cal C}_B^{{\rm TT}({\rm s})} \sim R_2/80$, 
the same
as for the EE ratio. Thus the upper limits correspond to an allowed
CMB contamination of this subdominant component of only $\sim
3\%$. For EE+TE, $q_{iso} =0.16 \pm 0.21$ with an upper limit of 55\%,
similar to the template value.

\section{Conclusions}
\label{sec:sum}

In this paper we present the results from 2.5 years of dedicated
polarization-optimized measurements with the Cosmic Background Imager.  From
this data, we estimate the TT, TE, EE and BB CMB angular power spectra.  The
EE power spectrum gives us a 11.7$\sigma$ detection of polarization, the
strongest thus far, while TE is measured at 4.2$\sigma$ versus zero.  The BB
spectrum gives a 95\% confidence upper limit of $3.8\, \mu\rm{K}^2$.

We introduce a novel method for the reconstruction of $\ell$-space maps of
$\tilde{E}$ and $\tilde{B}$.  Images of the E and B fields on the sky 
are formed by Fourier transform of the $\ell$-space maps; this is a new
way of representing CMB polarization and is complementary to the
standard Stokes Q and U images shown previously in \cite{paper8}.
The E-mode detection and the lack of one in B is evident in both the
raw maps and the reconstructed Wiener-filtered images of the $20^h$
strip and is also evident in the square mosaic fields. We have
also verified that signal-map fluctuations, shown in
Fig.~\ref{fig:deconvolved_20hr_real}(c,d), about the mean signal in 
Fig.~\ref{fig:deconvolved_20hr_real}(b) do not obscure
this clear detection: the $20^h$ strip is indeed dominated by the CMB
polarization signal. The signal maps of the total intensity, an
example of which is shown in Fig~\ref{fig:deconvolved_02hr_t} for the
$02^h$ mosaic, also show very strong detections.

An analysis of a six-dimensional space of cosmological parameters
shows that the patterns and amplitudes in the EE, TE and BB data are
entirely consistent with the basic inflation-based model predictions
from TT, a result considerably strengthened by the new CBI EE+TE
data. The combined CBI+DASI+B03 EE+TE data further sharpens this
conclusion.  This is particularly evident in
Fig.~\ref{fig:theta_model_plot} which shows that $\theta/\theta_0$,
parameterizing the angular scale associated with sound crossing at
decoupling and hence the peak-dip pattern, is pinned down to the value
we obtain from TT alone.

We finally explore a restricted physically-motivated class of
models with combined, but uncorrelated, adiabatic and isocurvature
perturbations.  We find that there is effectively no evidence for an
isocurvature mode in the data. Furthermore the data rule 
out a possible family of interloper peaks which would be out of phase
with the standard flat adiabatic predictions. This strengthens our
claim that cosmological models with an additional isocurvature 
mode are disfavored by the current polarization data.

\acknowledgements

We are indebted to the DASI team, of the Kavli Institute for Cosmological
Physics at the University of Chicago, led by John Carlstrom, and
especially to John Kovac, who designed the achromatic polarizers.
We thank the Kavli Operating Institute, Barbara and Stanley Rawn, Jr.,
Maxine and Ronald Linde, Cecil and Sally Drinkward, Rochus Vogt, and
the Provost, President, and PMA Division Chairman of the California
Institute of Technology for their generous support.  We gratefully
acknowledge support from the Canadian Institute for Advanced Research,
the Canadian Space Agency, and NSERC at CITA. This research used the
MacKenzie cluster at CITA, funded by the Canada Foundation for
Innovation.  This work was supported by the National Science
Foundation under grants AST 9413935, 9802989, 0098734, and 0206416.
RB was supported partially by CONICYT.  LB and JM acknowledge support
from the Chilean {\it Center for Astrophysics} FONDAP No.~15010003 and
ST from grant Milenio ICM P02-049.  This work was supported by the
Leverhulme Trust and PPARC at Oxford. ACT acknowledges support from
the Royal Society.  We thank CONICYT for granting permission to
operate within the Chanjnantor Scientific Preserve in Chile, and the
National Radio Astronomy Observatory (NRAO) Central Development Lab
for developing the HEMT amplifiers used in this project and assisting
with production.  The National Radio Astronomy Observatory is a
facility of the National Science Foundation operated under cooperative
agreement by Associated Universities, Inc. We thank Jo Dunkley, Carrie
MacTavish, and Mike Nolta for helpful comments.  We thank Nolberto
Oyarce, Wilson Araya, and Jos{\'e} Cortes for their dedicated work in
operating the CBI.

\bibliographystyle{apj}
\bibliography{ms.bib}

\end{document}